\documentclass{svjour3}                     % onecolumn (standard format)

\usepackage[top=30truemm,bottom=30truemm,left=25truemm,right=25truemm]{geometry}
\usepackage{lscape}
\usepackage[utf8]{inputenc}
\usepackage{url}

\usepackage{amsmath,amssymb,amsfonts}

\graphicspath{{Figures/}}
\usepackage{psfrag} % Must compile using TeX and DVI!!!  (Fig. labels)
\usepackage{graphicx} 
\usepackage{subcaption}
\usepackage[sort&compress,numbers]{natbib}
\usepackage[noend]{algpseudocode}
\usepackage{algorithm}
\usepackage{algorithmicx}
\newcommand*\Let[2]{\State #1 $\gets$ #2}
\algnewcommand\algorithmicinput{\textbf{Input:}}
\algnewcommand\Input{\item[\algorithmicinput]}
\algnewcommand\algorithmicoutput{\textbf{Output:}}
\algnewcommand\Output{\item[\algorithmicoutput]}
\algnewcommand\algorithmicinitialize{\textbf{Initialize:}}
\algnewcommand\Initialize{\item[\algorithmicinitialize]}
\algnewcommand\algorithmicoptimize{\textbf{Optimize:}}
\algnewcommand\Optimize{\item[\algorithmicoptimize]}
\renewcommand*\Call[2]{\texttt{#1}(#2)}

\newcommand{\matr}[1]{\mathbf{#1}} 
\newcommand{\vect}[1]{\mathbf{#1}}
\DeclareMathOperator{\sechsq}{sech^2}
\DeclareMathOperator{\sech}{sech}
\DeclareMathOperator{\prox}{prox}
\DeclareMathOperator{\proj}{proj}

\DeclareMathOperator*{\argmin}{argmin}

\begin{document}
\title{Dimensionality Reduction and Reduced Order Modeling\\ for Traveling Wave Physics
\thanks{We acknowledge the support from the Defense Threat Reduction Agency HDTRA1-18-1-0038, The Boeing Company, and the Army Research Office (W911NF-19-1-0045). }
}

\author{Ariana Mendible \and Steven L.~Brunton  \and Aleksandr~Y.~Aravkin \and Wes Lowrie \and J. Nathan Kutz}

\institute{A. Mendible \at
	Department of Mechanical Engineering, 
	University of Washington, Seattle, WA 98195, USA\\
	\email{mendible@uw.edu}
	\and
	S. L. Brunton\at
	Department of Mechanical Engineering, 
	University of Washington, Seattle, WA 98195, USA
	\and
	A. Y. Aravkin \at
	Department of Applied Mathematics,
	University of Washington, Seattle, WA 98195, USA
	\and
	W. Lowrie \at
	ARA
	\and 
	J. N. Kutz \at
	Department of Applied Mathematics,
	University of Washington, Seattle, WA 98195, USA
}

\date{Received: \today / Accepted: April 5, 2020}
\maketitle

	% ------------- ABSTRACT ------------- %
\begin{abstract}
	We develop an unsupervised machine learning algorithm for the automated discovery and identification of traveling waves in spatio-temporal systems governed by partial differential equations (PDEs).  Our method uses sparse regression and subspace clustering to robustly identify translational invariances that can be leveraged to build improved reduced order models (ROMs).  Invariances, whether translational or rotational, are well known to compromise the ability of ROMs to produce accurate and/or low-rank representations of the spatio-temporal dynamics.  However, by discovering translations in a principled way, data can be shifted into a coordinate systems where quality, low-dimensional ROMs can be constructed. This approach can be used on either numerical or experimental data with or without knowledge of the governing equations. We demonstrate our method on a variety of PDEs of increasing difficulty, taken from the field of fluid dynamics, showing the efficacy and robustness of the proposed approach.
\end{abstract}
	
\section{Introduction}

Spatio-temporal dynamical systems, characterized by partial differential equations (PDEs), are ubiquitous across science and engineering and are central to many fields such as fluid dynamics, neuroscience, and atmospheric science.  PDEs are rarely amenable to analytic, closed-form solutions, thus requiring recourse to computational solutions.  However, numerical discretization is prone to generating high-dimensional representations of the solution in order to accurately resolve multiple time and space scales and underlying nonlinearities~\cite{Kutz:2013}.  This is in contrast with the observation that the underlying dynamics often exhibit low-dimensional structure~\cite{HLBR_turb}.  
This is especially true in fluid systems which manifest many coherent structures dynamics such as vortices, large scale eddies, and traveling waves~\cite{HLBR_turb, duraisamy2019arfm}.
Reduced-order models (ROMs) exploit the intrinsic, low-rank structure of the simulation data in order to create more tractable models for the spatio-temporal evolution dynamics of the PDE. Many ROMs leverage the {\em singular value decomposition} (SVD) to produce a linear dimensionality reduction~\cite{Kutz:2013,Brunton2019book}, whereby a dominant set of correlated modes provide a subspace in which to project the PDE dynamics~\cite{Benner2015siamreview,Taira2017aiaa,taira2019aiaa}.   Typically low-energy modes are truncated, and the governing equations are projected onto the remaining high-energy modes to create an approximate and low-dimensional model. Dimensionality reduction and modal decomposition approaches have been well-studied and are extremely efficient~\cite{antoulas2005approximation,Benner2015siamreview,hesthaven2016certified,quarteroni2015reduced,Taira2017aiaa,taira2019aiaa}. The resulting models are computationally tractable, enabling downstream tasks such as prediction, estimation and low-latency control. However, many dimensionality reduction techniques are known to have severe limitations in capturing symmetry in the data, specifically translational symmetries found in traveling waves~\cite{Brunton2019book}.  In this paper, we develop an unsupervised, data-driven framework for identifying interpretable representations for traveling wave speeds that can be exploited for improved reduced-order models. 

ROMs have evolved into a critically enabling computational framework for PDE systems that are prohibitively expensive or intractable to simulate.  
Historically, ROMs arose in the fluid dynamics community in order to study the low-dimensional structures which are canonical to complex flows~\cite{HLBR_turb}.  Even high-dimensional turbulence can be more efficiently represented and analyzed within such a reduction framework.  Indeed, the majority of ROM simulations today for time-dependent PDEs are focused on fluid dynamic simulations, typically at high Reynolds numbers where improved simulation architectures are required for Monte-Carlo simulations for parametric studies and uncertainty quantification.
ROMs are tractable since they exploit the underlying low-dimensional nature of the spatio-temporal evolution dynamics.  Although emerging techniques in machine learning are being used to identify more effective coordinate systems and models for systems in fluid mechanics~\cite{Milano2002jcp,ling2016jfm, maulik2017jfm,Loiseau2018jfm,Brenner2019prf,Brunton2020arfm}, the majority of the time, low-rank structure used in ROMs is extracted using 
the {\em proper orthogonal decomposition} (POD)~\cite{Sirovich:1987,berkooz1993proper,HLBR_turb,Towne2018jfm}, which is typically computed via the SVD. If the governing equations are available, Galerkin projection~\cite{Sirovich:1987,berkooz1993proper,HLBR_turb,Noack2011book}, or the improved Petrov-Galerkin projection~\cite{carlberg2011efficient,carlberg2017galerkin},  can then be used to create an interpretable model of the few modes that contain the majority of the system's energy (variance). While POD-Galerkin projections have been quite successful in many cases~\cite{Noack2003jfm,tadmor2011reduced}, this approach requires access to the governing equations, and there are well-known issues with the POD subspace~\cite{morzynski2007continuous}.  In addition to POD-based methods, data-driven methods for ROMs have gained traction, such as the {\em dynamic mode decomposition} (DMD)~\cite{Rowley2009jfm, Schmid2010jfm, Tu2014jcd, Kutz2016book} which creates a best-fit linear operator to advance spatio-temporal snapshots forward in time. 

\begin{figure}[t]
	\centering
	\begin{subfigure}[t]{0.313\textwidth}
		\includegraphics[width=\textwidth]{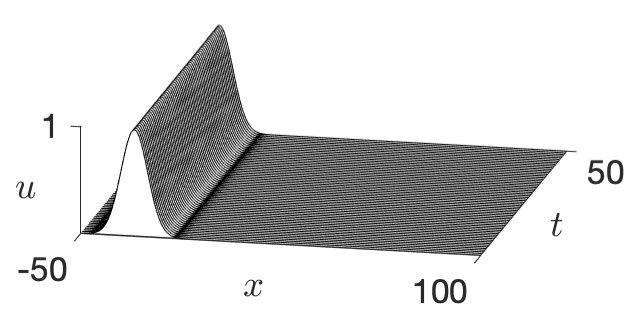}
		\subcaption{}\label{fig:stationary}
	\end{subfigure}
	\hfill
	\begin{subfigure}[t]{0.313\textwidth}
		\includegraphics[width=\textwidth]{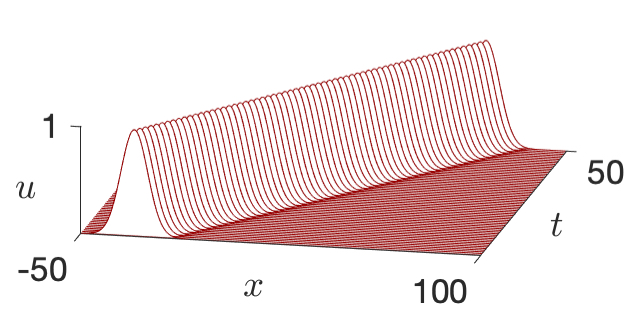}
		\subcaption{}\label{fig:traveling}
	\end{subfigure}
	\hfill
	\begin{subfigure}[t]{0.313\textwidth}
		\includegraphics[width=\textwidth]{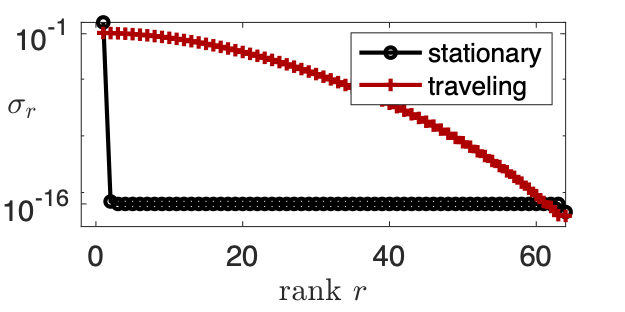}
		\subcaption{}\label{fig:podsvd}
	\end{subfigure}
	\caption{(\subref{fig:stationary}) Stationary soliton wave, (\subref{fig:traveling}), same soliton wave traveling in time, (\subref{fig:podsvd}) Singular value spectrum of the stationary wave (black circles) and the traveling wave (red pluses). The singular value $\sigma_r $ represents the energy contained by, or importance of, of the $r^\text{th}$ spatial mode. The singular values of the stationary wave show that the first mode contains all the energy, and other modes are zero to machine precision. Conversely, the traveling wave singular values decay slowly, indicating that many modes are needed to accurately reconstruct this wave. }
		\label{fig:podfailure}
\end{figure}

Regardless of the ROM architecture used, many PDE systems of importance contain symmetries that are obvious to the observer, such as rotations, translations, and scaling. Recognizing these symmetries makes object tracking and forward-time prediction trivial for a human, for example to estimate where and when a baseball will land from a limited set of observations. In traditional and widely-accepted ROM approaches, however, these symmetries greatly restrict our ability to produce meaningful dimensionality reduction.  Specifically, SVD-based methods are known to fail in handling rotations, translations, and scaling.  More precisely, the SVD is only appropriate for the small subset of dynamical systems where time and space interactions can be decoupled through separation of variables.  
In fluids, such symmetries and translations play a foundational role in the dynamics, such as the traveling waves observed in pipe flows~\cite{kerswell} or the Tollmien-Schlichting waves which arise   
from bounded shear flow, as in as boundary layers and channel flows~\cite{bagheri2009jfm}. 
For even moderately complex fluid flows the effective rank for POD is still too high because this tool is unable to handle these invariances \cite{taira2019aiaa, rowley2004physd, Reiss2018jsc, balajewicz2013jfm, bagheri2009jfm}.
Thus many systems are not amenable to POD/SVD model reduction strategies because this separability assumption fails to hold-- the space and time dynamics are nonlinear and inherently coupled. To illustrate this canonical phenomenon, the POD/SVD of a stationary and traveling soliton are compared in Figure \ref{fig:podfailure}. The stationary wave is perfectly captured by one spatial mode, as seen in the rapid singular value decay, whereas a wave translating in time cannot be represented by a low dimensional representation with POD/SVD. This space-time separation problem has received surprisingly little attention, but research is growing~\cite{kirby1992reconstructing,Rowley2000physd,Rim2018juq,Reiss2018jsc}. The focus of this paper will be the study of reduced-order modeling techniques for transport-dominated systems characterized by traveling waves.  Our unsupervised data-driven methods can be applied with or without knowledge of the governing equations, thus providing a robust mathematical architecture for ROMs exhibiting wave phenomenon.   

%%%%%%%%%%%%%%%
%%%%%%%%%%
\subsection{Background}

Previous work~\cite{Iollo2014pre,Lucia2001cfd, Cagniart2019book, Mojgani2017arxiv} has focused on using modal decomposition techniques, such as POD, in single transport phenomenon. Others have also explored symmetry groups and coordinate transforms \cite{Mojgani2017arxiv,Fedele2015jfm,Gerbeau2014jcp,Iollo2014pre} for decompositions, but these methods all require knowledge of the underlying equations, which are not always readily available. Instead, it is often desirable to focus on methods that will be effective on either simulated equations or experimental data with no closed-form solution.

Many approaches to creating ROMs for traveling waves have focused on using a template fitting approach as demonstrated by Kirby and Armbruster~\cite{kirby1992reconstructing} to shift waves into a transport-invariant coordinate frame. In this template fitting approach, the shift amount $c(t)$ is determined by projecting $u(x,t)$ onto a template $u_0(x)$, often taken to be the initial condition. This approach is well-suited to single waves which maintain a near-constant waveform, though multiple waves, breathing or dispersion characteristics cannot be effectively captured. In 2000, Rowley and Marsden~\cite{Rowley2000physd} developed a method of reconstructing systems with translational symmetry by pre-shifting the POD modes by a speed $c(t)$ discovered through template fitting with the initial condition, resulting in:
\begin{equation}\label{eqn:mrshift}
u(x,t) = \sum_{r= 1}^N a_r(t) \phi_r(x-c(t)). 
\end{equation}
While this approach may handle even complicated periodic systems well, such as the Kuramoto-Sivashinsky equation, the model creation is based inherently in the known dynamics of the system. This approach does not generalize well to systems with unknown governing equations, similar to methods developed by others in~\citep{Mojgani2017arxiv, Fedele2015jfm, Gerbeau2014jcp, Iollo2014pre}. Work by Iollo and Lombardi~\cite{Iollo2014pre} accounted for diffusive phenomenon in traveling wave decomposition by relying on the governing equations to separate advective and diffusive components. Without relying on equations of motion, Lucia et al~\cite{Lucia2001cfd} were able to create reduced-order models for shock waves by separating out and reducing the transport-dominated domain. This approach may introduce an undesirable bias in the selection of the domains to decompose, and here we aim for an unsupervised approach.  

Rim et al~\cite{Rim2018juq} uses the template fitting approach to determine the so-called shift numbers in which the wave is stationary. They perform a transport reversal to shift each time step of the snapshot matrix into the moving coordinate frame of the transport phenomenon. They also propose solutions to many of the difficulties introduced by a periodic shift operator and quick evolution of the transported structures. They present a scaling vector to account for changing wave heights, a cut-off map for non-periodic boundaries, regularization for a smooth and interpretable transport vector, and a greedy algorithm to account for multiple or rapidly-changing transported quantities. The authors show examples of the linear wave equation for demonstration, Burgers' equation to show shock formation, and the acoustic equation with heterogeneous materials to exhibit variable wave speeds. Here, we aim to build \textit{interpretable} models for the shift speeds, which will allow understanding of the physics and future state of the system. Nonetheless, these innovations are promising computational tools that should be incorporated into future work. 

Most closely related to this work is the work of Reiss et al~\cite{Reiss2018jsc}, who introduced a data-driven decomposition for multiple traveling waves. The {\em shifted Proper Orthogonal Decomposition} (sPOD) views the snapshot matrix $u(x_i,t_n)$ of discretized spatiotemporal data in $N_s$ moving coordinate frames, decomposing each with POD:
\begin{equation}\label{eqn:spod}
u(x_i,t_n) \approx \sum_{k=1}^{N_s} T^{c_k} \left( \sum_r \alpha_r^k(t_n)\phi_r^k (x_i) \right). 
\end{equation}
The sPOD algorithm uses the discrete shift operator $T^{c_k}$, corresponding to the transport speeds $c_k$, to pre-shift the data. After aligning data into a moving coordinate frame, POD is iteratively used to determine the modes for each transport speed until a prescribed error tolerance is met. The shift operator is found in one of two ways: using a thresholding algorithm or using an SVD-based search in a prescribed interval. In the thresholding method, the threshold must be chosen ad hoc, and a potentially noisy or non-smooth set of shift values is returned. In the SVD-based shift search, many candidate constant shift values $c$ must be supplied, and the SVD computation is performed for each in order to determine which leads to the maximum leading singular values (or quickest decaying singular value spectrum). This leads to a potentially high computational cost and a lack of robustness to non-constant wave speeds. In addition, both of these supervised methods yield black box shift values that are not robust to drastically evolving wave shapes or non-periodic boundaries.  

Regardless of these limitations, Reiss et al~\cite{Reiss2018jsc} provides an exceptional step forward for handling traveling waves in PDEs.  Building on this work, we propose a data-driven, optimization-based approach to utilize symmetries present in traveling waves for dimensionality reduction with interpretable physics. We explore the use of our speed detection method on four examples of traveling waves, varying in complexity. We demonstrate that our method efficiently accounts for non-periodicity and non-constant speeds in multiple interacting traveling waves, and uncovers the relevant wave speed physics. We then develop ROMs, comparing various reduction approaches. Finally, we present further improvements and future work.

\begin{figure}[tb]
	\includegraphics[width=\textwidth]{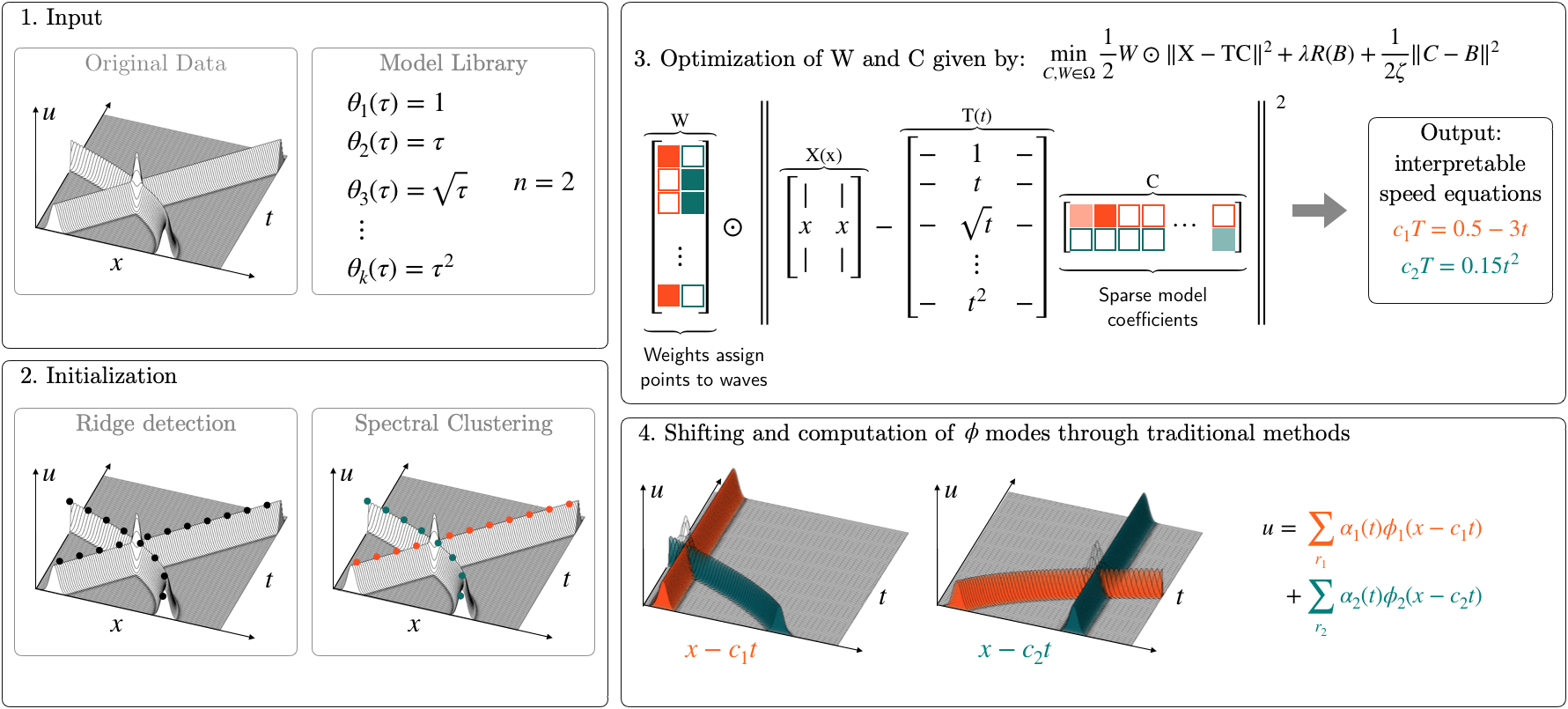}
	\caption{Schematic of the optimization-based shift mechanism. (1) The input data is passed in as a 2D snapshot matrix $\matr{U}$. A model library is constructed by choosing simple functions of time to describe candidate wave speeds, and a number of waves is chosen. (2) To initialize, a ridge detection is performed and finds the locations of the wave ridges as $(\vect{x},\vect{t})$ pairs. Then, spectral clustering is performed to preliminarily assign the $(\vect{x},\vect{t})$ pairs to each wave, shown separated into orange and blue waves. (3) An optimization is performed over the $\matr{W}$ and $\matr{C}$ matrices, with $\matr{B}$ acting as an auxiliary variable. $\matr{W}$ governs the assignment of the $(\vect{x},\vect{t})$ pairs to each wave. $\matr{C}$ gives the coefficients of the library terms that account for each wave speed, orange and blue. This matrix is penalized to be sparse so that only a few library terms are chosen out of many candidates. The speeds for each wave are interpreted as each row of the $\matr{C}$ matrix. (4) The original data is shifted into each wave's reference frame, making each a stationary wave. Traditional model reduction techniques can then be used to find $\phi$ modes for each. }
	\label{fig:schematic}
\end{figure}

\section{Unsupervised Traveling Wave Identification with Shifting and Truncation (UnTWIST)}

Our mathematical architecture provides a method for pre-shifting traveling wave data in order to improve the performance of standard dimensionality reduction techniques for ROMs. We develop an optimization framework for discovering interpretable wave speed models for multiple, non-constant speed waves, called UnTWIST ({\em Unsupervised Traveling Wave Identification with Shifting and Truncation}).   

\subsection{The Optimization Formulation }\label{subsec:wavedecomp}

For a traveling wave with time-varying speed $c(t)$, the function $u(x,t) = f(x-c(t))$ describes the waveform over space and time. The first task is to determine a set of $\vect{x}$ and $\vect{t}$ pairs of wave peak points that follow the path of the traveling wave over time. Ridge detection is a ubiquitous and well-studied task in computer vision~\cite{rasche2017rapid}, by which we can determine these points. The number of wave peak points $N$ is arbitrary, but a more refined discretization of space and time will yield higher $N$ and more accurate peak point detection.

 \paragraph{Simple Case.}
 In the case of a single traveling wave with constant speed $c$, we can formulate the optimization as  $\vect{x}-c\vect{t}=0$. Determining the scalar value of $c$ that best describes these points is a simple least-squares regression. In most cases of interest, however, wave speeds are not  constant. Rather, they may be non-constant, nonlinear, or have some prescribed time dynamics in the form of $\vect{x}-c(\vect{t})=0$, as seen in the spatial modes $\phi$ of \eqref{eqn:mrshift}.

\paragraph{Non-Constant Speeds.}
In order to describe waves with speed functions $c(\vect{t})$, we form a matrix $\matr{T}$ of $k$ candidate wave functions of time, for example, $\vect{1}$, $\vect{t}^2$ or $\sin(\vect{t})$, that may describe the wave's non-constant speed over time. The choice of library functions allows the user enormous flexibility to customize the method to their application. The functions can be thoughtfully chosen based on visualization or similarity to other known dynamical systems, though this is not necessary. Using a rich library of simple terms such as polynomial and trigonometric functions usually yield strong candidate models. The input variable $n$ gives the number of waves present in the wave field. We repeat our $\vect{x}$ vector $n$ times to form the $\matr{X}$ matrix. Searching for a wave's speed is a least-squares problem which yields \textit{coefficients} $\matr{C} \in \mathbb{R}^{k \times n}$ to potential wave speed functions:
\begin{equation}\label{eq:xminusct}
	\min_\matr{C} \frac{1}{2} \|\matr{X-TC}\|^2_2 .
\end{equation}
The matrix $\matr{C}$ is composed of rows corresponding to each term in the candidate function library $\matr{T}$. Columns of $\matr{C}$ correspond to each wave.  In general, we will seek the simplest model for the wave speed, given by the fewest terms or the sparsest matrix $\matr{C}$, similar to the sparse identification of nonlinear dynamics (SINDy)~\cite{Brunton2016pnas}.

\paragraph{Multiple Waves.}
When more than one wave is present in a wave field, however, a mask must be used to differentiate between the wave speed model for each wave. The weighting matrix $\matr{W}$ is implemented to isolate rows of $\matr{X}$ and $\matr{T}$, corresponding to wave peak points, which are active for each wave. Here, $\matr{W}\in\mathbb{R}^{N \times n}$ assigns $N$ points to $n$ waves, 
with each row of $\matr{W}$ encoding a normalized probability distribution for each wave peak point. Entries are forced toward binary values of 0 or 1, corresponding to a wave peak point belonging to only one wave with high probability.
We formulate the optimization to include multiple waves and non-constant wave speeds as
\begin{equation}
	\min_{\matr{C}, \matr{W} \in \Omega} \frac{1}{2} \matr{W}  \odot \|\matr{X-TC}\|^2_2,
\end{equation}
where the symbol $\odot$ denotes the Hadamard element-wise product. 

\paragraph{Parsimonious Models.}
Ultimately, the success of this method relies on accommodating a wide variety and arbitrary number of candidate functions to express the proper wave speed. However, only a small subset of these candidate functions will be relevant for characterizing a real system; even the most complex physically-based systems, such as Navier-Stokes, contain relatively few terms. With this in mind, we introduce a \textit{sparsity} promoting constraint, usually in the form of a $\ell_0$- or $\ell_1$-norm~\cite{Brunton2016pnas}. This additional constraint penalizes a dense matrix and forces many terms to zero. We apply this sparsity constraint on the $\matr{C}$ matrix so that many of the coefficients of the candidate model functions become zero, i.e. the wave speeds have few active terms in the library. If this sparsity constraint were not imposed, the $\matr{C}$ matrix generically contains many high-order terms since this minimizes a least-square $\ell_2$-norm. This is seldom, if ever, physically meaningful. Further, it generally leads to instability for future-time prediction, and is prone to over-fitting. 

\subsection{Sparse Relaxed Regularized Regression Optimization}\label{subsec:sr3}
Enforcing a sparsity constraint on the wave speed models makes the optimization challenging, and in some cases, nonconvex. While many methods have been developed to address the pursuit of sparsity, such as the ubiquitous LASSO~\cite{tibshirani1996regression}, % more citations? 
{\em Sparse Relaxed Regularized Regression} (SR3)~\cite{zheng2018ieee} was chosen because of its adaptability to different nonconvex sparsity constraints and its computational efficiency.
The SR3 formulation of the optimization problem to fit speed models to multiple waves with non-constant candidate functions is formulated as
\begin{equation}
	\label{eq:mainsr3}
	\min_{\matr{C}, \matr{B}, \matr{W} \in \Omega} \frac{1}{2} \matr{W \odot \|X-TC}\|^2_2 + \lambda R(\matr{B}) + \frac{1}{2\zeta} \|\matr{C-B}\|^2_2.
\end{equation}
The regularizing function $R(\matr{B})$ is chosen to be a sparsity constraint, for example the $\ell_0$- or $\ell_1$-norm is used so that  $R(\matr{B})=\|\matr{B}\|_0$, for instance. The number of nonzero terms is controlled by the value of the $\lambda$ parameter. The matrix $\matr{B}$ is an auxiliary variable which relaxes the sparsity constraints by being enforced here rather than on $\matr{C}$ directly. $\matr{B}$ is forced to be close to $\matr{C}$, with tuning from the relaxation parameter $\zeta$. Tuning the hyperparameters $\lambda$ and $\zeta$ carefully is critical to reaching a meaningful minimum, although there is no {\em a priori} knowledge of the optimal values. Generally, $\lambda$ is considered the sparsity parameter and should be chosen such that only handful of terms in the library are allowed to remain active. The $\zeta$ parameter can be considered a step size and should be chosen such that the optimization changes significantly enough in error for each iteration, but that it does not take too many iterations to converge.  In the examples presented, a judicious choice of hyperparameters was made for the sake of simplicity. However, this could be done in a principled fashion by using a parameter sweep, and choosing optimal values. Hyperparameter tuning is standard in most machine learning algorithms~\cite{witten2016data}. The SR3 optimization architecture has proven accurate, robust, and efficient in nonconvex problems with sparsity constraints such as the one presented.

\subsection{Implementation}\label{subsec:implementation}

\noindent Algorithms~\ref{alg:untwist} details the UnTWIST algorithm.  Algorithm~\ref{alg:sr3} details the corresponding sparsity promoting SR3 algorithm used for selecting a wave speed model.

\begin{algorithm}[h!]
	\caption{UnTWIST}
	\label{alg:untwist}
	\begin{algorithmic}[1]
		\Input{$\matr{U}$: wave field, $x$ and $t$: space-time discretization, $n$: number of waves, $\boldsymbol{\theta}_{1:k}$: library of candidate speed models, and $\lambda$ and $\zeta$: optimization hyperparameters}
		\Output{A} 
	
		\Initialize{}
			\Let{$(\vect{x},\vect{t})$}{ \Call{RidgeDetection}{$\matr{U}$} } \Comment{Find wave peak points, see Figure~\ref{fig:schematic}.2}
			\Let{$\matr{T}$}{$\boldsymbol{\theta}_{1:k}(\vect{t})$} \Comment{Form library from candidate speed functions, from Figure~\ref{fig:schematic}.1}
			\Let{$\matr{X}$}{\Call{repmat}{$\vect{x},k$}} 
			\Let{$\matr{W}_0$}{\Call{SpectralClustering}{$\vect{x}, \vect{t}, n$} } \Comment{Find probable wave groups, see Figure~\ref{fig:schematic}.2}
			% could add for loop for each wave finding a speed function using TLS
			\Let{$\matr{C}_0$}{\Call{ThresholdedLeastSquares}{$\matr{W}_0, \vect{x}, \vect{t}$} } \Comment{Initialize models for each wave speed}
%		\Optimize{}
			\Let{($\matr{B}_{1:\nu}, \matr{C}_{1:\nu}, \matr{W}_{1:\nu}$)}{\Call{SR3}{$\matr{C}_0,\matr{W}_0$}} \Comment{Optimize using SR3, detailed in Algorithm~\ref{alg:sr3}, see Figure~\ref{fig:schematic}.3}
		\Let{$\matr{U}_{1:n}$}{\Call{Shift}{$n, \matr{U}, \matr{C}_\nu, \matr{T}$}} \Comment{Shift in each reference frame, see Figure~\ref{fig:schematic}.4}
		\Let{$\sum_{i=1}^r{\alpha_i \phi(x)_i}$}{\Call{POD}{$\matr{U}_{1:n}$}}
	\Comment{Any dimensionality reduction method may be used, see Figure~\ref{fig:schematic}.4}
	\end{algorithmic}
\end{algorithm}

\begin{algorithm}[h!]
	\caption{SR3 for UnTWIST}
	\label{alg:sr3}
	\begin{algorithmic}[1]
		\State {\bfseries Input:} $\matr{C}_0, \matr{W}_0, \lambda, \zeta$
		\State {\bfseries Initialize:} $\nu=0, \matr{B}_0 = \matr{C}_0$
		\While{not converged}
		\Let{$\nu$}{$\nu+1$}
		{\Let{$\matr{C}_{\nu+1}$}{$\argmin_{\matr{C}}\left(\frac{1}{2} \matr{W} \odot \|\matr{X}-\matr{T}\matr{C}\|^2 + \frac{1}{2\zeta} \|\matr{C}-\matr{B}\|^2 \right) $}}
		{\Let{$\matr{B}_{\nu+1}$}{$\prox_{\lambda \eta R}\left(\matr{C}_{\nu+1}\right) $}}		
		\Let{$\matr{W}_{\nu+1}$}{$\proj_{\Omega} \left( (\matr{X} - \matr{T}\matr{C}_{\nu+1}) \odot (\matr{X}-\matr{T}\matr{C}_{\nu+1}) \right) $}
		\EndWhile
		\State {\bfseries Output:} $\nu, \matr{C}_{1:\nu}, \matr{B}_{1:\nu}, \matr{W}_{1:\nu} $
	\end{algorithmic}
\end{algorithm}

It is important to note that initialization of our algorithm plays a key role in the success of this method over a very broad optimization space. To initialize the weighting matrix $\matr{W}$, as seen in a clustering performed on the initial $\vec{x}$ and $\vec{t}$ data, assigning initial probabilities that each wave peak point belongs to a given wave. We employed spectral clustering~\cite{ng2002spectral} to obtain these initial probabilities, relying on the eigenvalue spectrum of a similarity matrix of the points. %The symmetric adjacency matrix was filled with ones wherever two points are ``neighbors" within a given radius of one another. The Laplacian matrix subtracts the diagonal degree distribution, or total number of ``neighbors" for a given point, from this matrix. Then, a k-means clustering was performed on the first eigenvector of the matrix. 
This yields points that are connected along a spectrum, rather than within a simple Euclidean distance from a centroid, as with many other clustering algorithms. Points in a given cluster are assigned a weight of unity in a column of $\matr{W}_0$ and zero elsewhere, with each column of $\matr{W}_0$ corresponding to a different cluster, or wave. 

The $\matr{C}$ matrix of speed function coefficients was initialized using a sequential thresholded least-squares algorithm, as in~\cite{Brunton2016pnas}. % ~\cite{Brunton2016pnas}.
In this scheme, the least-squares solve is performed for each row of the $\matr{C}$ matrix, corresponding to each wave. The smallest terms will be thresholded out and ``deactivated" to enforce sparsity, another least-squares is performed, and the process is repeated until the sparsity constraint is satisfied. This yields a sparse initial model for each wave as coefficients in $\matr{C}_0$. 

The SR3 optimization is comparable in computational cost to the ADMM algorithm, a popular and ubiquitous optimization scheme, with a one-time overhead of $\mathcal{O}(Nn^2+n^3)$ and iteration cost of $\mathcal{O}(n^2)$. It is important to note that here, this cost does not scale with the size of the original data. Rather, it scales with the number of points selected by the ridge detection, i.e. the size of the $\matr{X}$ and $\matr{T}$ matrices, which is typically much smaller than the input. In future use for higher dimensional systems, this cost would not scale proportional to the additional dimensions but with the same points denoting the translated quantity, which would be higher but on the order of magnitude of the 1D examples. One-time computations in the initialization include 10 total operations on the original data for the ridge detection \cite{rasche2017rapid}, a $k$-means clustering, which scales with $\mathcal{O}(Nn)$.

The optimization given by \eqref{eq:mainsr3} yields the $\matr{C}$ matrix, which is used with the library $\matr{T}$ to determine the interpretable speed models for each of the waves. %Interpretable speed models allow one to discover underlying patterns and physics from data alone. It can also be used to forecast wave speeds and extrapolate in forward-time to make predictions using ROMs. 
These speed models are used to shift each time slice of the input data matrix. This effectively places the wave field into a traveling reference frame in which one wave appears stationary. Traditional decompositions such as POD are then used and result in meaningful and low-rank modes for ROM reconstruction.

\section{Results}

We demonstrate the UnTWIST algorithm on four different linear and nonlinear fluid dynamic systems that exhibit traveling waves. We first show the initialization of the data from the spectral clustering and model selection steps. We present the outcome of the UnTWIST algorithm as well as performance metrics. Errors are computed as the $\ell_2$-norm of the difference between discovered wave speed model coefficients and the known wave speed model coefficients taken from the underlying equations. Finally, we present the dimensionality reduction resulting from various methods using the UnTWIST algorithm. 
It is important to note that the choice of the model library and hyperparameters are critical to the convergence of the optimization and the parsimony of the speed models. Each set of parameters is included for the corresponding examples. Additionally, code for all examples is provided on Github at \url{https://github.com/mendible/wave_decomposition}.

\subsection{UnTWIST Results}

\paragraph{Example 1: Korteweg-de Vries Equation.}
The first example is one of the simplest examples of a traveling wave: a single soliton of constant speed propagating over a domain with periodic boundary conditions. The evolution of the wave field governed by the Korteweg-de Vries (KdV) equation is given by 
\begin{equation}\label{eq:kdv}
u_t+uu_x+u_{xxx}=0
\end{equation}
with initial condition
\begin{equation}\label{eq:kdv_ic}
u_0(x) = 1+24\sechsq{(\sqrt{2}x)}.
\end{equation}

\begin{figure}[tb]
	\begin{centering}
		\includegraphics[width=0.7\textwidth]{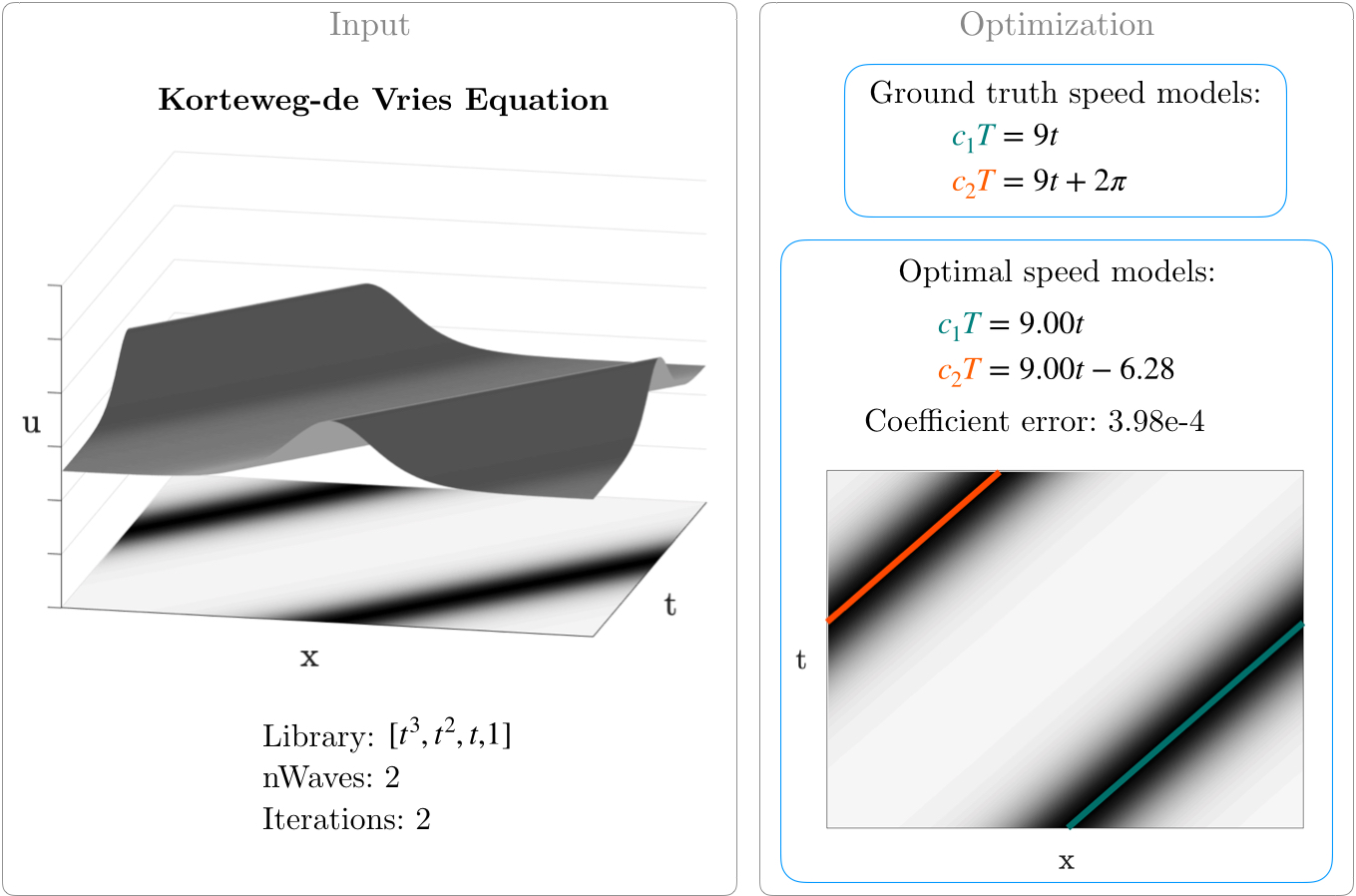}
		\caption{Results of the UnTWIST algorithm on the simulated KdV equation input data. The algorithm converged to a coefficient  error of $3.98e{-4}$ within 2 iterations, and is close to the ground truth speed used in simulation.}
		\label{fig:kdv_results}
	\end{centering}
\end{figure}

The KdV equation is widely used to describe gravity waves propagating through shallow water~\cite{ablowitz2011nonlinear}.
The initial condition \eqref{eq:kdv_ic} is a known soliton solution of the KdV equation, which maintains a constant wave shape, with a wave propagation speed of 9.  % weird to not have units here
This initial condition was discretized into 512 grid points in $[-\pi,\pi)$ and solved for $t \in [0, 0.6]$ using an exponential time-differencing fourth-order Runge-Kutta scheme~\cite{kassam2005fourth}, where results were saved at steps of $\Delta t =0.001$ in time. The initial data can be seen in the surface plot in Figure~\ref{fig:kdv_results} (left panel).

A library of potential wave speeds, as given in Figure~\ref{fig:kdv_results}, was chosen to be polynomials of the time variable up to third order, with a constant included for centering the data in $x$. 
The number of waves was chosen to be $n=2$ in order to allow the algorithm to converge to a meaningful and interpretable wave speed model. A single wave model could have been used, and in this case, found to be a parallel wave directly in between the periodic extension of the right-traveling wave. Hyperparameters were tuned to be $\lambda = 10$ and $\zeta = 1e{-4}$.

For this simple example, where the wave speeds are constant in time, the UnTWIST algorithm performed with high accuracy, yielding coefficients within $3.98e{-4}$ of the ground truth wave speed models, written explicitly in Figure~\ref{fig:kdv_results}. The two wave crests were identified clearly with the ridge detection step. The algorithm converged in only two iterations, indicating that the initialization via spectral clustering and thresholded least squares solved the optimization nearly to tolerance, and only one optimization step was taken. Figure~\ref{fig:kdv_results} also shows that the wave speed models, given in orange and green lines, closely follow the wave crests.

\begin{figure}[tb]
	\begin{centering}
		\includegraphics[width=\textwidth]{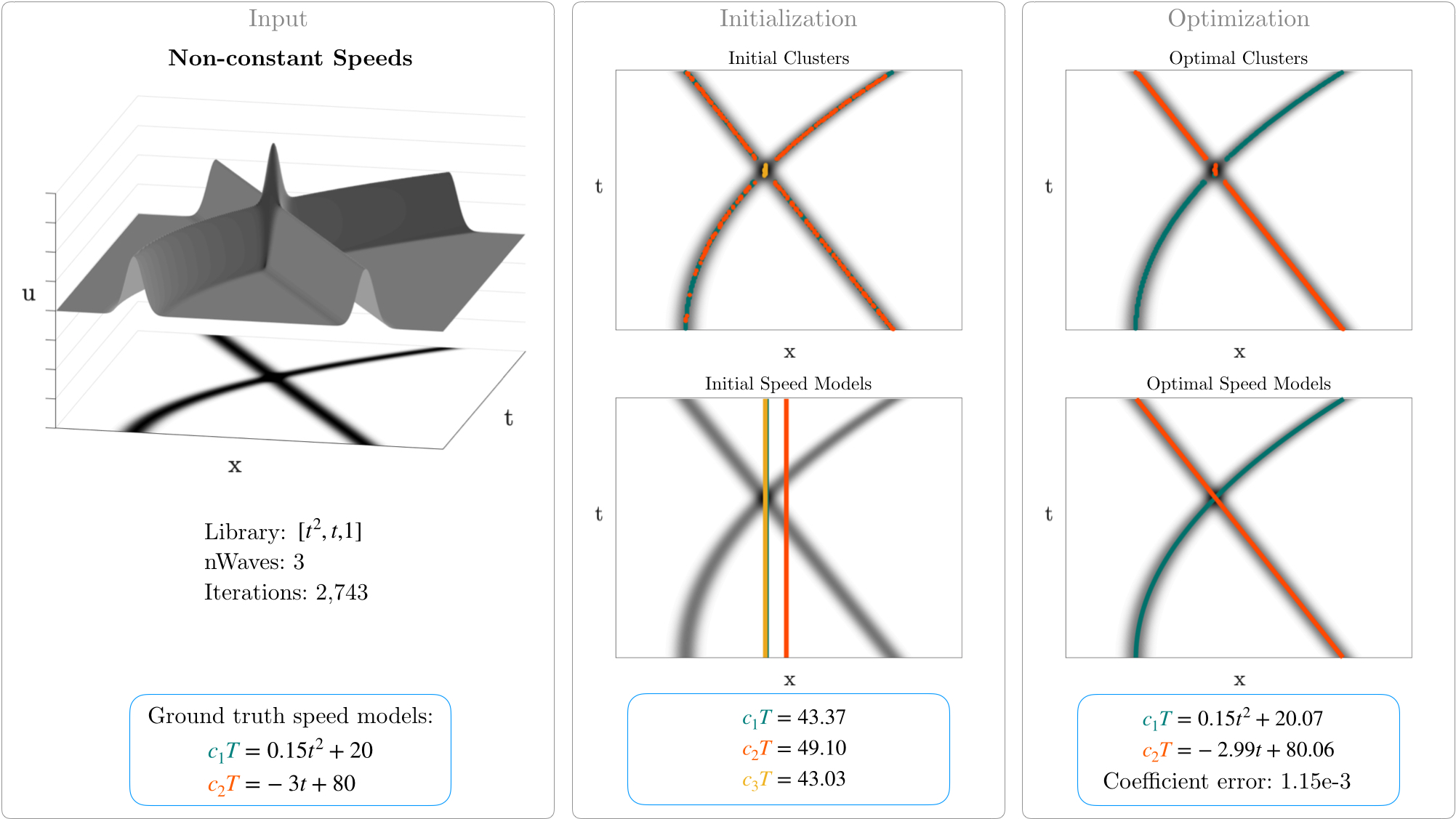}
		\caption{Results of the optimization on artificially generated data of non-constant velocity linear soliton waves and a library as given.}
		\label{fig:complicated_results}
	\end{centering}
\end{figure}

\paragraph{Example 2: Two Crossing Solitons: Non-periodic and Non-Constant Speed.}
The second example we present is a fabricated non-periodic wave field of two crossing Gaussian solitons of non-constant speed. The wavefield was constructed using the governing equation
\begin{equation}\label{eq:complicated}
u(x,t) = \exp{\left(-0.1(x+3t-80)^2\right)}+\exp{\left(-0.1(x-0.15t^2-20)^2\right)}
\end{equation}
on $x, t \in [0,100] \times [0,20]$ in a uniform $(256 \times 512)$ grid. The library is taken to be polynomial terms of $t$ up to second-order. Hyperparameters were tuned to be $\lambda = 10,000$ and $\zeta = 1e{-4}$. The number of waves $n$ was chosen to be 3 as an initial value. The UnTWIST algorithm allows the number of waves to change if the number of points in a cluster shrink below a threshold in the optimization step. In this example, giving a starting value of $n=3$ allowed a subset of points to be in a temporary third wave group until a more probable wave belonging is found. It can be seen in Figure \ref{fig:complicated_results} that the addition of a third wave group allows the yellow points, corresponding to the complicated wave intersection area, to belong to a wave group of their own. After the optimization converged in 2,743 iterations, the third wave group has fewer points than the threshold allows, so this model is removed. These points are assigned to the right-traveling wave, and the models closely fit the wave crests. The speed models also have close agreement, with the coefficient error at $1.15e{-3}$ compared to the ground truth speeds.

\begin{figure}[tb]
	\begin{centering}
		\includegraphics[width=0.7\textwidth]{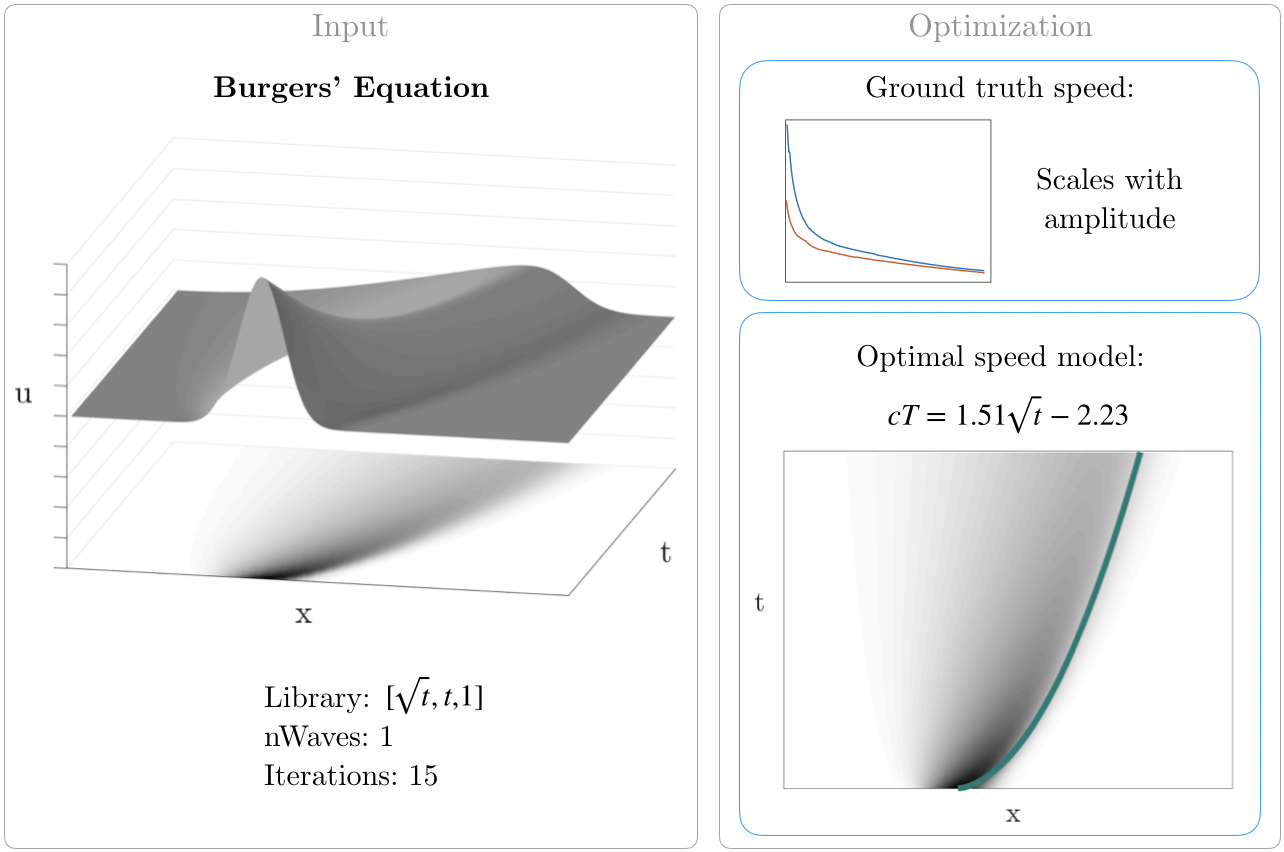}
		\caption{Results of the UnTWIST algorithm on data from Burgers' equation. Speeds were determined and compared to the height of the wave as determined by ridge detection, and follow the same trend.}
		\label{fig:burgers_results}
	\end{centering}
\end{figure}

\paragraph{Example 3: Burgers' Equation.}
The third example we present uses viscous Burgers' Equation, \eqref{eq:burgers}. Because of its characteristic of forming sharp discontinuities, viscous Burgers' equation is used to describe shock waves common in gas dynamics~\cite{ablowitz2011nonlinear}. This provides an example of a changing wave profile alongside a non-constant wave propagation speed, as seen in Figure \ref{fig:burgers_results}. This example does not provide a known speed, but it is well-known that the wave front propagates at a speed proportional to the wave height. % reference?

\begin{equation}\label{eq:burgers}
u_t + uu_x - 0.1u_{xx} =0 .
\end{equation}
This equation was solved using fourth-order Runge-Kutta scheme for $t\in [0,20]$ in steps of $\Delta t = 0.1$ in a periodic domain of $x \in [-8,8)$ with 256 grid points. The initial condition is given in \eqref{eq:burgers_ic}, representing a Gaussian wave.
\begin{equation}\label{eq:burgers_ic}
u_0(x)=\exp{(-(x+2)^2)}.
\end{equation}

The library was chosen to be a linear, square-root, and constant term. The square-root term was chosen in order to accommodate the nonlinear speed simply. The number of waves was chosen to be $n=1$. Hyperparameters were tuned to be $\lambda = 10$ and $\zeta = 1e{-4}$. As seen in Figure \ref{fig:burgers_results}, the wave speed model determined by the optimization fits closely to the data, converging in only 15 iterations. Since there is no known ground truth, the wave speed model was also compared to the wave amplitude over time, as seen in Figure \ref{fig:burgers_results}. The close agreement indicates that the wave speed was well-identified. This example also illustrates the success of the ridge detection algorithm for this application, as wake peak points are captured well, even with changing wave height. 

\begin{figure}[tb]
	\begin{centering}
		\includegraphics[width=\textwidth]{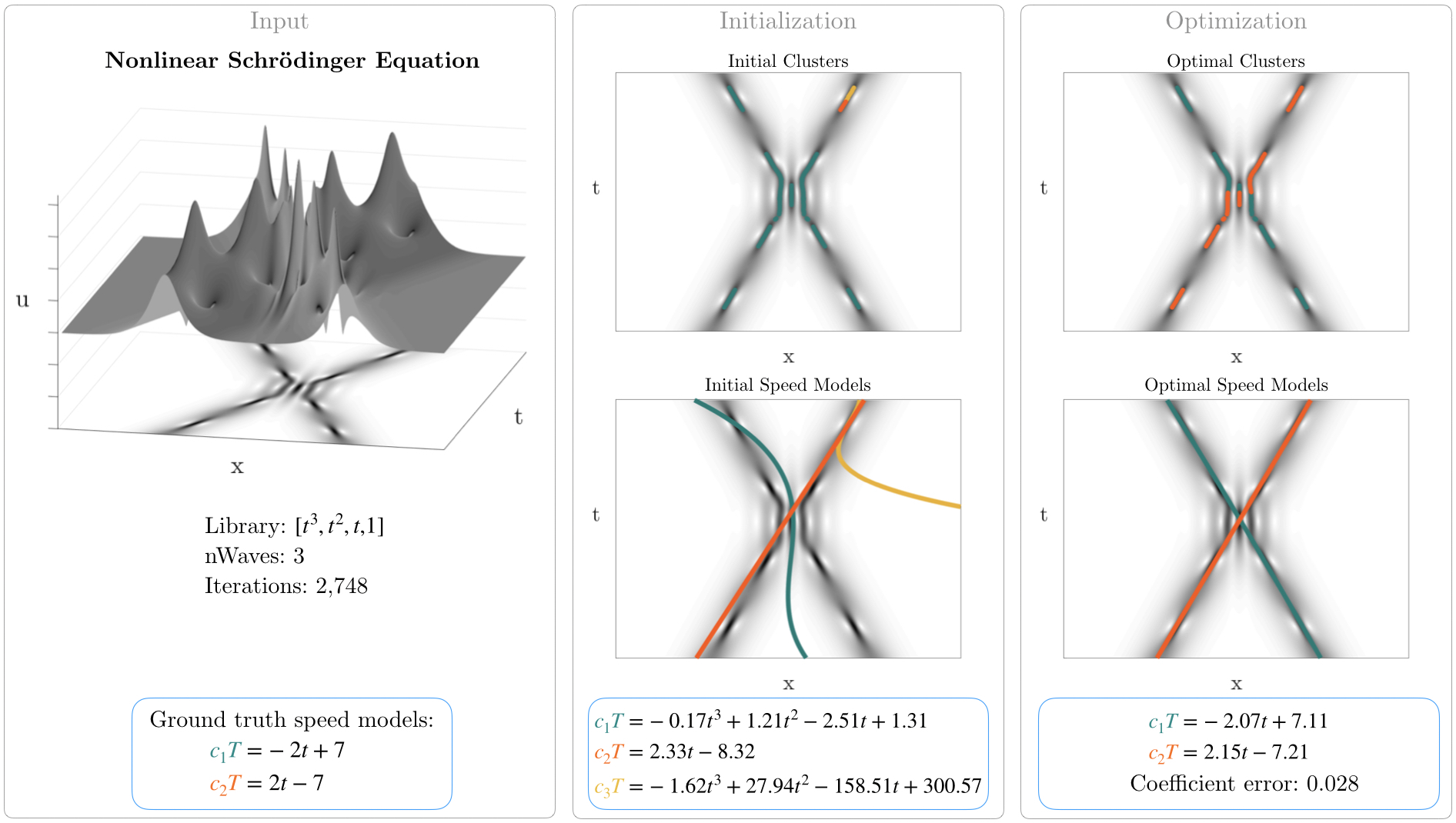}
		\caption{Results of the UnTWIST algorithm to find the shifts in simulated NLS equation data. Three clusters are used to allow the algorithm to threshold out complicated areas at first. The initial clustering is poor, as are corresponding initial models. However, after the optimization, the clustering has improved, and the models are visually an excellent fit. The extraneous model is ignored and the error in the coefficients is 0.028 after 2,748 iterations.}
		\label{fig:nls_results}
	\end{centering}
\end{figure}

\paragraph{Example 4: Nonlinear Schrödinger  Equation.}
The final example we present showcases many of the complications we wish to address with the UnTWIST algorithm-- nonlinear wave interaction, non-periodicity, and rapidly-changing wave profiles. The complex nature of the wave field can be seen in Figure \ref{fig:nls_results}. The Nonlinear Schrödinger Equation is used across many fields. In fluid mechanics, it describes gravity-driven surface water waves in the deep water regime~\cite{ablowitz2011nonlinear}, and is typically solved using expensive and finely-tuned numerical methods. NLS is given as 

\begin{equation}\label{eq:nls}
iu_t = -\frac{1}{2}u_{xx}+\kappa|u|^2u. 
\end{equation}
The 
This was solved on the domain $x,t \in [-15,15)\times[0,2\pi]$ in a $1024 \times 501$ uniform grid. The initial condition 
\begin{equation}\label{eq:nls_ic}
u_0=2\sech{(x+7)}\exp{(2ix)}+2\sech{(x-7)}\exp{(-2ix)}
\end{equation}
was propagated forward in time using a pseudo-spectral, fourth-order Runge-Kutta scheme~\cite{Kutz:2013}.

For this example, the library of candidate functions was made up of polynomials up to third order including a constant term. Similar to the fabricated data in Example 3, this wave field has complicated wave crossings, so an extra wave group was initialized with $n=3$ waves. Hyperparameters were tuned to be $\lambda = 10$ and $\zeta = 1e{-4}$. As seen in Figure \ref{fig:nls_results}, the initialization did not cluster the points into a centered wave-crossing section as expected, rather, they are scattered. The initial models also show very poor correlation to the wave crests. After the optimization with 2,748 iterations, however, the models show a very close agreement to the wave crests and the extraneous wave cluster is eliminated. The discovered wave speed models also show close agreement with a coefficient error of 0.028 compared to the ground truth speeds. 

\subsection{Reduced Order Models}
We now present low-rank reconstructions of two of the example wave fields using UnTWIST. We couple our algorithm with two common dimensionality reduction methods, POD and Robust Principal Component Analysis (RPCA)~\cite{Candes:2011}. We compare these rank-reduced wave field reconstructions with those of the traditional (unshifted) POD approach, each using only 2 modes. The singular value spectrum is also given for each example, showing the amount of information contained by each successive mode in the decomposition. It is important to note that when shifting is applied, we obtain $n$ wave fields. For the shifted reconstructions, dimensionality reduction is performed for each of resulting pre-shifted wave fields. The first mode is taken from each shifted wave field and all are superimposed, each counting as one mode when comparing to traditional POD. 

\begin{figure}[tb]
\centering
\begin{subfigure}[b]{0.45\textwidth}
	\includegraphics[width=\textwidth]{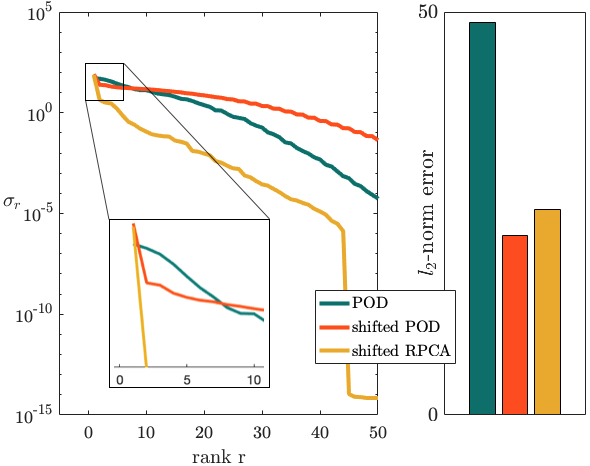}
	\caption{Singular value spectra and error at rank 2}
	\label{fig:complicated_spec}
\end{subfigure}
\begin{subfigure}[b]{0.45 \textwidth}
	\includegraphics[width=\textwidth]{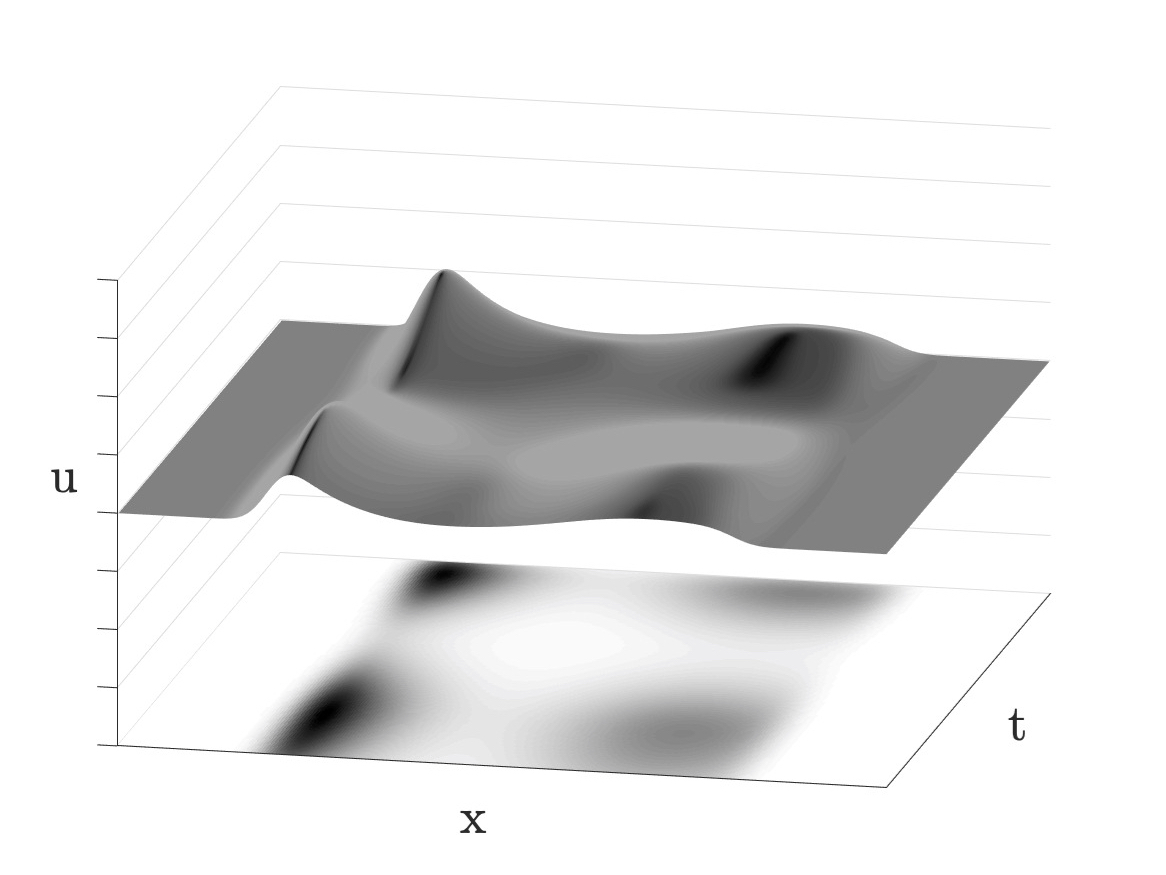}
	\caption{Traditional POD}
	\label{fig:complicated_pod}
\end{subfigure}
\begin{subfigure}[b]{0.45\textwidth}
	\includegraphics[width=\textwidth]{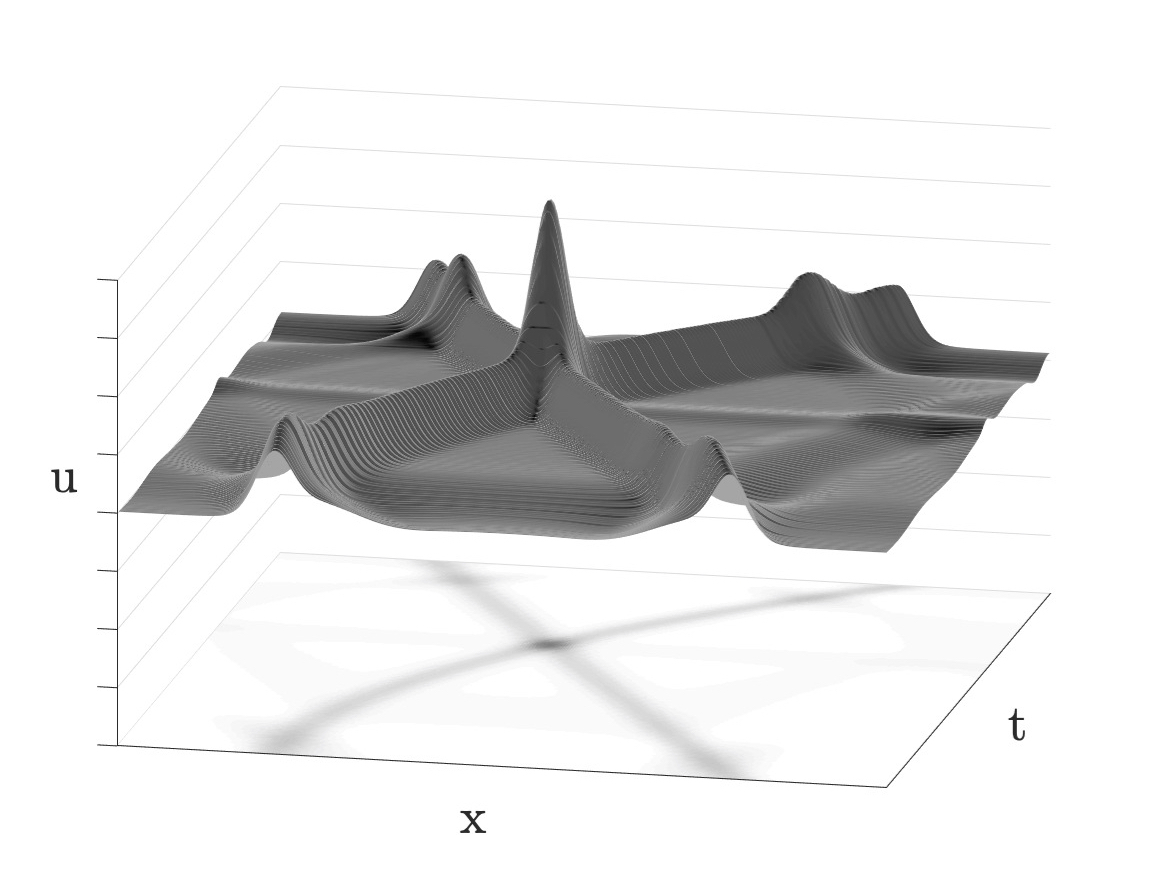}
	\caption{Shifted POD}
	\label{fig:complicated_spod}
\end{subfigure}
\begin{subfigure}[b]{0.45\textwidth}
	\includegraphics[width=\textwidth]{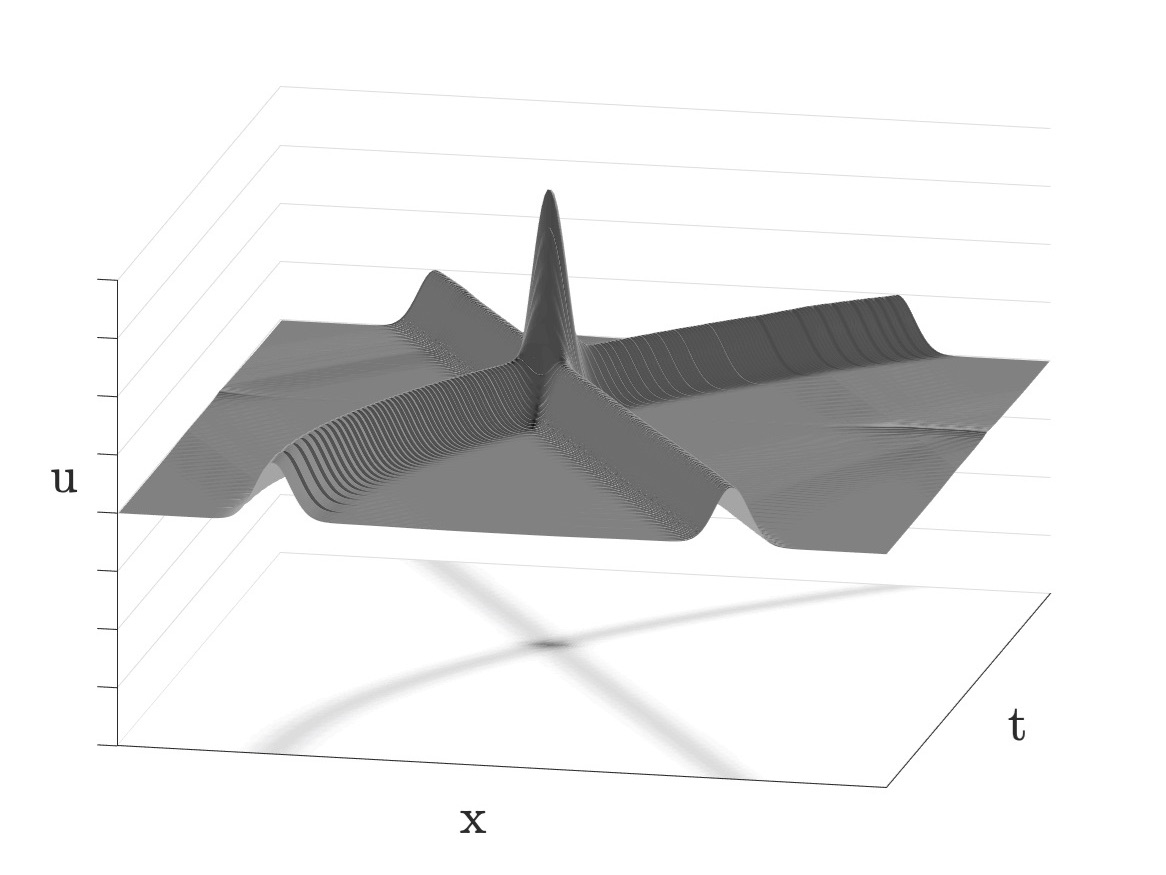}
	\caption{Shifted RPCA}
	\label{fig:complicated_srpca}
\end{subfigure}
\caption{Comparison of different dimensionality reduction methods for generated example data, each using 2 modes. \ref{fig:complicated_spec} shows the singular value spectra for each dimensionality reduction method for various ranks, showcasing the energy captured in each successive rank. Error plots at right show the $\ell_2$-norm error for each method with a rank-2 reconstruction, whose surface plots are shown in \ref{fig:complicated_pod}-\ref{fig:complicated_srpca}.}\label{fig:complicated_roms}
\end{figure}

The first example wave field is described in Example 2, presenting complications of non-constant wave speeds, non-periodicity, and multiple crossing waves. As seen above, the wave speeds are well-identified using the UnTWIST method. The data was pre-shifted into two frames using the two identified speed functions and both POD and RPCA were performed. These are presented in Figure \ref{fig:complicated_roms} and compared to traditional POD with all methods truncating to rank-2.

The singular value spectra for each of these decompositions is show in Figure \ref{fig:complicated_roms}. It can be seen that the shifted RPCA captures the most energy out of the three methods out to around 75 modes, or about 30\% of the original size. In this case, traditional POD appears to perform preferably to the shifted POD, when viewing the total spectrum. It can be seen that for 2-8 modes of retention, shifted methods both outperform traditional POD. In the images of the reconstructions, the two-mode reconstruction is clearly superior in terms of faithfully representing the dynamics of the system. The shifted RPCA performs the best, with two clear solitons interacting and maintaining smoothness outside this domain. The shifted POD somewhat recreates the wave pattern, but with artifacts obscuring the smooth soliton behavior. In the traditional POD example, however, the wave field is unrecognizable. There is no continuity between the two solitons and there is no clear traveling behavior represented.

\begin{figure}[tb]
	\centering
	\begin{subfigure}[b]{0.45\textwidth}
		\includegraphics[width=\textwidth]{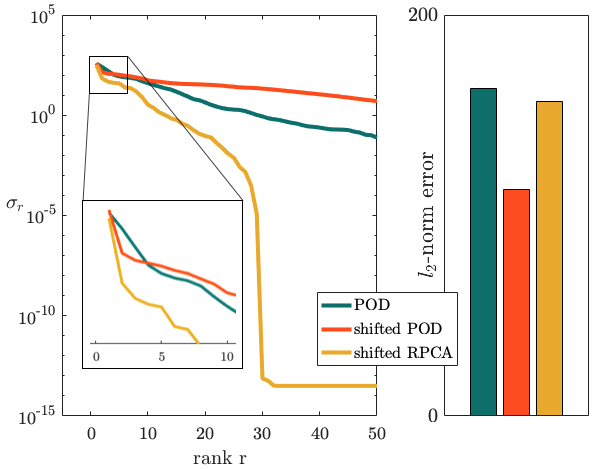}
		\caption{Singular value spectra and error at rank 2}
		\label{fig:nls_spec}
	\end{subfigure}
	\begin{subfigure}[b]{0.45 \textwidth}
		\includegraphics[width=\textwidth]{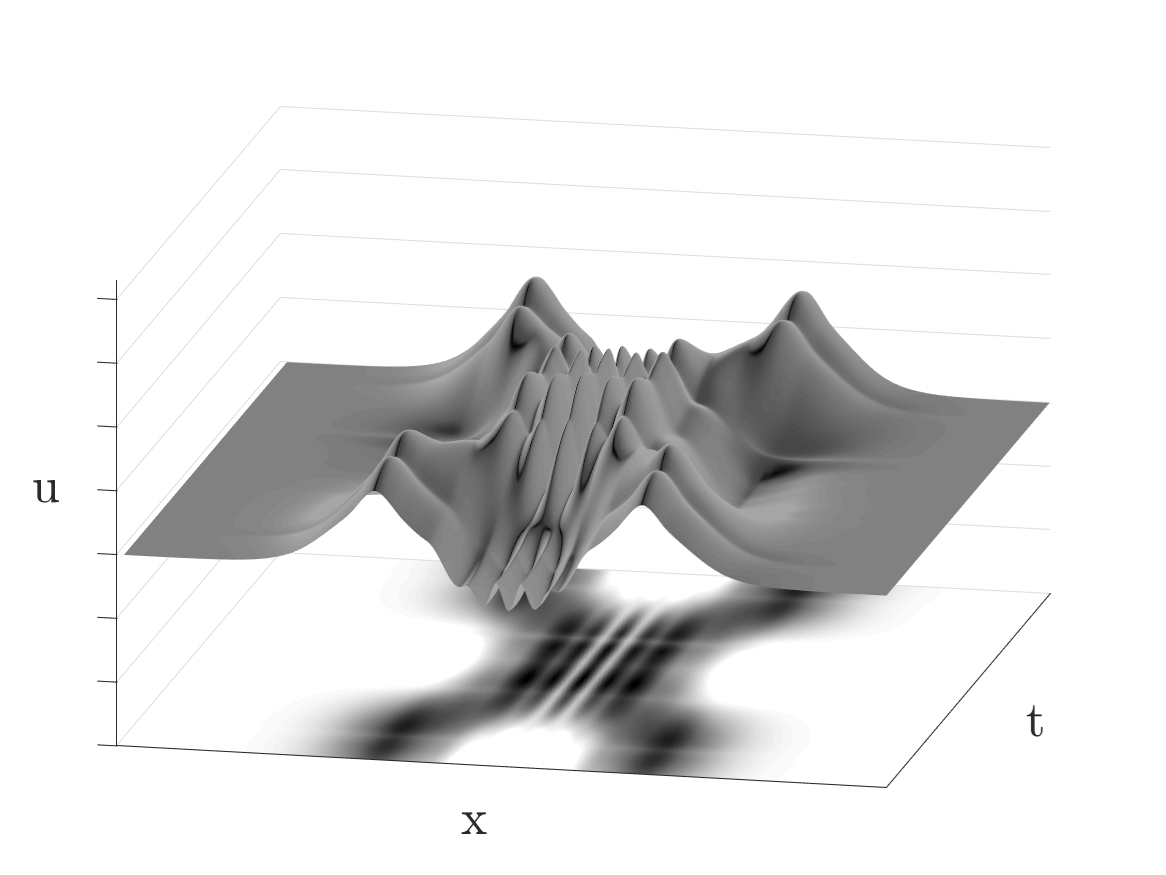}
		\caption{Traditional POD}
		\label{fig:nls_pod}
	\end{subfigure}
	\begin{subfigure}[b]{0.45\textwidth}
		\includegraphics[width=\textwidth]{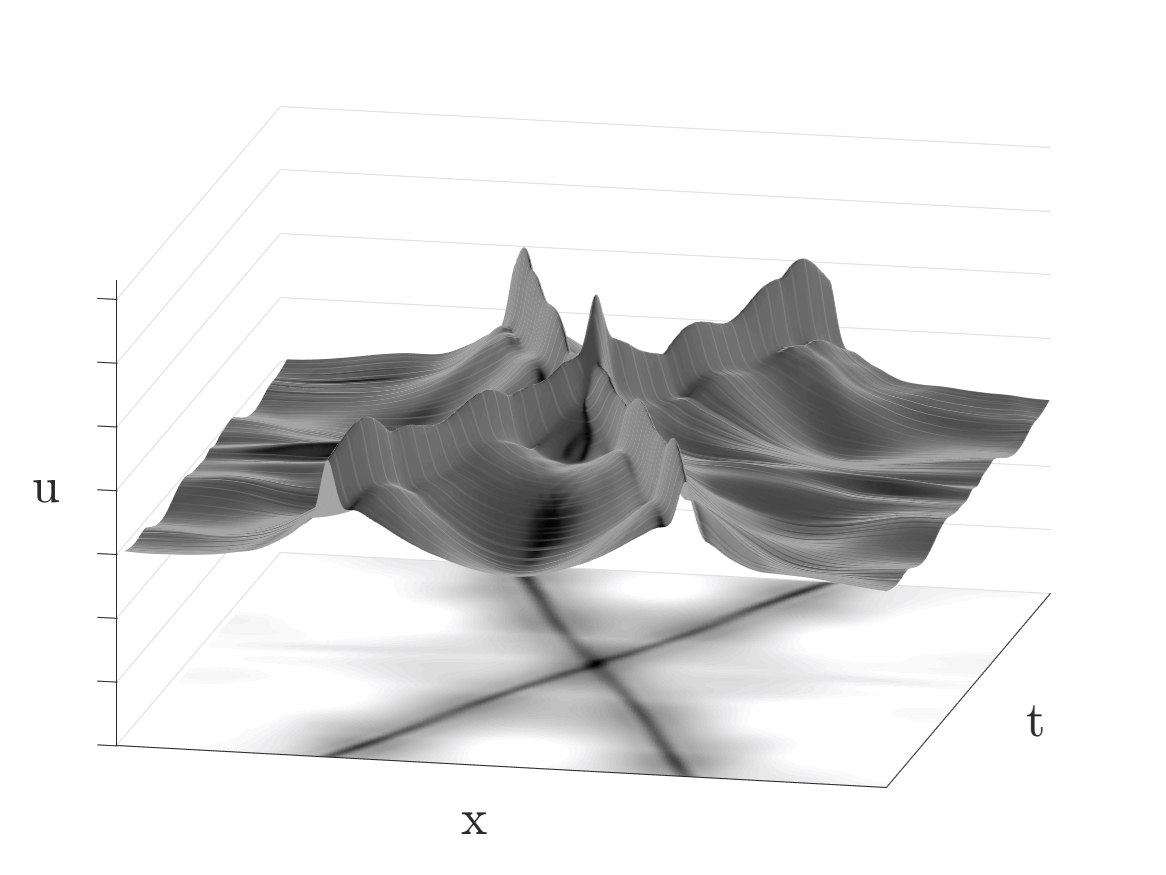}
		\caption{Shifted POD}
		\label{fig:nls_spod}
	\end{subfigure}
	\begin{subfigure}[b]{0.45\textwidth}
	\includegraphics[width=\textwidth]{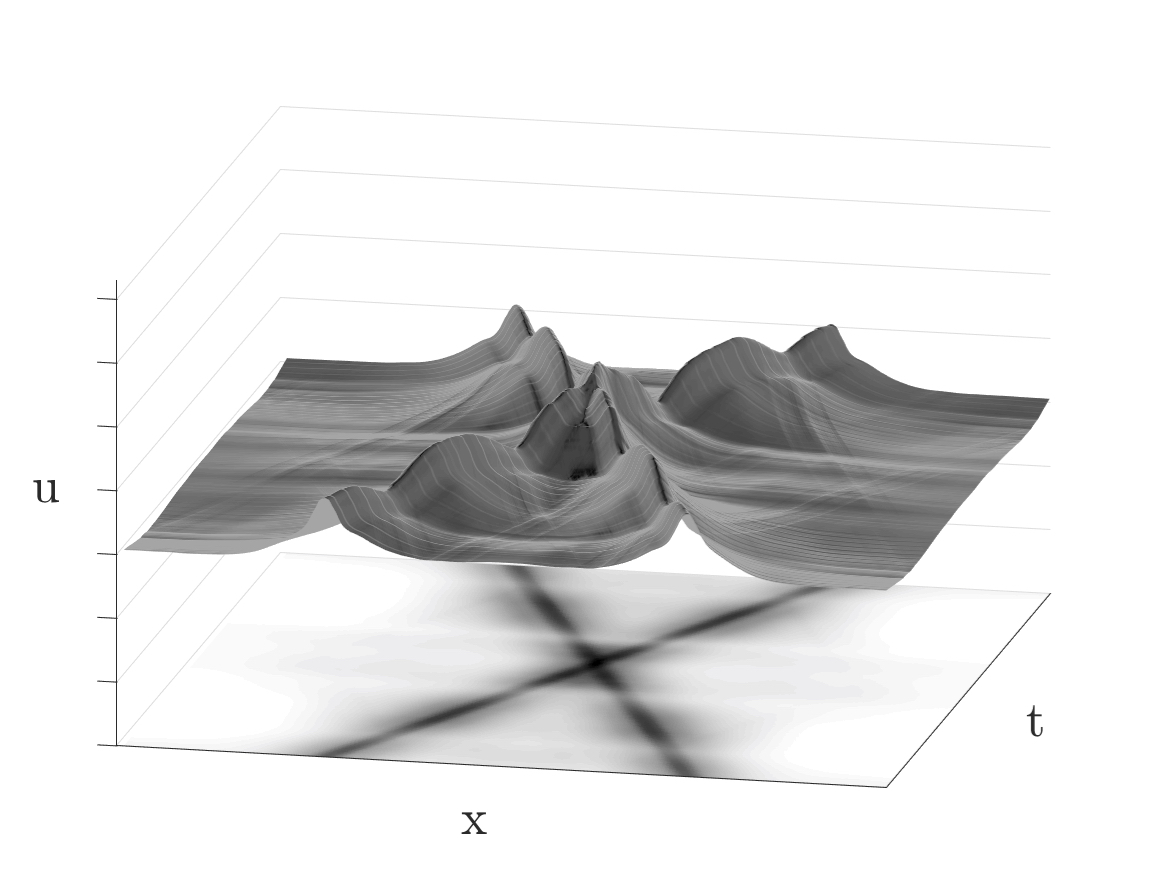}
	\caption{Shifted RPCA}
	\label{fig:nls_srpca}
\end{subfigure}
	\caption{Comparison of different dimensionality reduction methods for NLS data, each using 2 modes. \ref{fig:nls_spec} shows the singular value spectra for each dimensionality reduction method for various ranks, showcasing the energy captured in each successive rank. Error plots at right show the $\ell_2$-norm error for each method with a rank-2 reconstruction, whose surface plots are shown in \ref{fig:nls_pod}-\ref{fig:nls_srpca}. }\label{fig:nls_roms}
\end{figure}

The second example we present is of Example 4, the NLS equation. The results of the decompositions are shown in Figure~\ref{fig:nls_roms}. It can be seen in the singular value spectrum that for this case, the shifted RPCA decomposition outperforms the others for all ranks of truncation. The shifted POD again outperforms the traditional POD for few modes, from 2-4 modes retained. The traditional POD quickly takes over the shifted, however, and captures more information for 5 and higher modes than the shifted version. Overall, it can be seen clearly that the shifted RPCA quickly captures the most possibly information in only about 35, or about 7\% of all possible, modes. The rank-2 reconstructions illustrate this concept similarly. While the shifted RPCA does not have the characteristic sharp peaks of the original data, the area outside the breathers is relatively smooth. In contrast, both the shifted POD and the traditional POD maintain oscillations near the edges of the breathers. These are a product of the many high frequency oscillatory modes needed to represent the breathing waves.

%should this be in the Disicussion/conclusion/improvements?
The success of these shifted reconstructions may hinge on the ability of the dimensionality reduction algorithm to filter background from higher frequency phenomenon. When implementing a shift into a moving reference frame, a single traveling wave is viewed as stationary in time. When more than one wave is present in the field, however, shifting simply distorts the wave's speed. This problem is akin to a background separation-- one low-frequency, stationary wave is the background of wave field, and the other is the foreground. When performing a dimensionality reduction, this mode is often the most energetic. Many dimensionality reduction methods, including RPCA, have focused on isolating these background modes, explaining the success of RPCA in this example. 

\section{Conclusions and Improvements}
% context of work
Dimensionality reduction methods play a key role in reduced order modeling across a broad range of engineering applications. In this paper, we have shown the difficulty of using traditional methods such as POD on systems with translational symmetry in the data, namely, traveling waves. We then presented an unsupervised machine learning method for identifying interpretable representations for the speeds of these traveling waves. Using the speeds, it is possible to pre-shift the data before applying traditional dimensionality reduction methods in order to reduce the number of modes needed to accurately represent the data.

% set apart the work and reiterate the method-- connect innovations to needs in the field
While other approaches have similarly aimed to improved traditional methods, this work expands the application of pre-shifting methods to systems whose physics and dynamics are less well-known. The unsupervised nature of this method allows for an unbiased dimensionality reduction, regardless of whether the equations are known \textit{a priori}. Whether data is generated from experiments or numerical simulation of equations, the UnTWIST algorithm can uncover translational symmetries. The interpretable nature of the translations also allows one to uncover dynamics within the data. This proves especially important when there is little prior understanding of the underlying physics.

% drawbacks/challenges of this method and improvements in the near term
The primary improvement that can be made to the UnTWIST algorithm is including a term for the wave field reconstruction in the objective function. The way the optimization is currently formulated, each wave speed model is fit independently, and they are not considered as a superposition in the original data. Primarily, this leads to issues in the reconstruction when multiple waves are considered -- superimposing two waves can skew the intersection points to be double the height. In the future, assigning wave heights optimally to match the original data, as is done in~\cite{Reiss2018jsc}, is necessary to creating accurate reconstructions.  Indeed, our unsupervised learning algorithm for the detection of waves with the algorithms of~\cite{Reiss2018jsc} provides an excellent architecture for ROM efforts.

% future steps in the longer term
Despite the challenges in implementing UnTWIST thus far, the opportunities to expand its applications are promising. In this work, we have presented a method of identifying \textit{translational} symmetries in the data. In future work, unsupervised identification of rotational symmetries will allow for more efficient dimensionality reduction of a broader class of problems. Many fluids systems of interest, including flow across a cavity or over an airfoil, present vortex shedding that is not easily captured in few modes with traditional methods. Implementing a method to identify rotational symmetries and pre-rotate data may lead to considerable savings in simulation time for fluid systems.
Furthermore, the opportunities of this method to identify arbitrary symmetries are exciting. Given any symmetry group parameterized by $c = f(x,t)$, unsupervised machine learning methods are well-suited to uncover underlying patterns that may not be obvious to the observer. Expanding methods such as UnTWIST to allow for arbitrary symmetry identification is a promising path to significant improvements in dimensionality reduction.

\section*{Acknowledgements} 
We acknowledge the support from the Defense Threat Reduction Agency (DTRA) HDTRA1-18-1-0038 and the Army Research Office (W911NF-19-1-0045).

%\newpage

\bibliography{references.bib}{}

\begin{thebibliography}{10}

\bibitem{Kutz:2013}
J.~N. Kutz.
\newblock {\em Data-Driven Modeling \& Scientific Computation: Methods for
  Complex Systems \& Big Data}.
\newblock Oxford University Press, 2013.

\bibitem{HLBR_turb}
P.~J. Holmes, J.~L. Lumley, G.~Berkooz, and C.~W. Rowley.
\newblock {\em Turbulence, coherent structures, dynamical systems and
  symmetry}.
\newblock Cambridge Monographs in Mechanics. Cambridge University Press,
  Cambridge, England, 2nd edition, 2012.

\bibitem{duraisamy2019arfm}
Karthik Duraisamy, Gianluca Iaccarino, and Heng Xiao.
\newblock Turbulence modeling in the age of data.
\newblock {\em Annual Review of Fluid Mechanics}, 51:357--377, 2019.

\bibitem{Brunton2019book}
S.~L. Brunton and J.~N. Kutz.
\newblock {\em Data-Driven Science and Engineering: Machine Learning, Dynamical
  Systems, and Control}.
\newblock Cambridge University Press, 2019.

\bibitem{Benner2015siamreview}
Peter Benner, Serkan Gugercin, and Karen Willcox.
\newblock A survey of projection-based model reduction methods for parametric
  dynamical systems.
\newblock {\em SIAM review}, 57(4):483--531, 2015.

\bibitem{Taira2017aiaa}
Kunihiko Taira, Steven~L Brunton, Scott Dawson, Clarence~W Rowley, Tim
  Colonius, Beverley~J McKeon, Oliver~T Schmidt, Stanislav Gordeyev, Vassilios
  Theofilis, and Lawrence~S Ukeiley.
\newblock Modal analysis of fluid flows: An overview.
\newblock {\em AIAA Journal}, 55(12):4013--4041, 2017.

\bibitem{taira2019aiaa}
Kunihiko Taira, Maziar~S Hemati, Steven~L Brunton, Yiyang Sun, Karthik
  Duraisamy, Shervin Bagheri, Scott Dawson, and Chi-An Yeh.
\newblock Modal analysis of fluid flows: Applications and outlook.
\newblock {\em arXiv preprint arXiv:1903.05750}, 2019.

\bibitem{antoulas2005approximation}
Athanasios~C Antoulas.
\newblock {\em Approximation of large-scale dynamical systems}, volume~6.
\newblock Siam, 2005.

\bibitem{hesthaven2016certified}
Jan~S Hesthaven, Gianluigi Rozza, Benjamin Stamm, et~al.
\newblock {\em Certified reduced basis methods for parametrized partial
  differential equations}.
\newblock Springer, 2016.

\bibitem{quarteroni2015reduced}
Alfio Quarteroni, Andrea Manzoni, and Federico Negri.
\newblock {\em Reduced basis methods for partial differential equations: an
  introduction}, volume~92.
\newblock Springer, 2015.

\bibitem{Milano2002jcp}
Michele Milano and Petros Koumoutsakos.
\newblock Neural network modeling for near wall turbulent flow.
\newblock {\em Journal of Computational Physics}, 182(1):1--26, 2002.

\bibitem{ling2016jfm}
Julia Ling, Andrew Kurzawski, and Jeremy Templeton.
\newblock Reynolds averaged turbulence modelling using deep neural networks
  with embedded invariance.
\newblock {\em Journal of Fluid Mechanics}, 807:155--166, 2016.

\bibitem{maulik2017jfm}
Romit Maulik and Omer San.
\newblock A neural network approach for the blind deconvolution of turbulent
  flows.
\newblock {\em Journal of Fluid Mechanics}, 831:151--181, 2017.

\bibitem{Loiseau2018jfm}
J.-C. Loiseau, B.~R. Noack, and S.~L. Brunton.
\newblock Sparse reduced-order modeling: sensor-based dynamics to full-state
  estimation.
\newblock {\em Journal of Fluid Mechanics}, 844:459--490, 2018.

\bibitem{Brenner2019prf}
MP~Brenner, JD~Eldredge, and JB~Freund.
\newblock Perspective on machine learning for advancing fluid mechanics.
\newblock {\em Physical Review Fluids}, 4(10):100501, 2019.

\bibitem{Brunton2020arfm}
Steven~L. Brunton, Bernd~R. Noack, and Petros Koumoutsakos.
\newblock Machine learning for fluid mechanics.
\newblock {\em to appear in Annual Review of Fluid Mechanics (arXiv preprint
  arXiv: 1905.11075)}, 52, 2020.

\bibitem{Sirovich:1987}
L.~Sirovich.
\newblock Turbulence and the dynamics of coherent structures, parts {I-III}.
\newblock {\em Q. Appl. Math.}, XLV(3):561--590, 1987.

\bibitem{berkooz1993proper}
Gal Berkooz, Philip Holmes, and John~L Lumley.
\newblock The proper orthogonal decomposition in the analysis of turbulent
  flows.
\newblock {\em Annual review of fluid mechanics}, 25(1):539--575, 1993.

\bibitem{Towne2018jfm}
Aaron Towne, Oliver~T Schmidt, and Tim Colonius.
\newblock Spectral proper orthogonal decomposition and its relationship to
  dynamic mode decomposition and resolvent analysis.
\newblock {\em Journal of Fluid Mechanics}, 847:821--867, 2018.

\bibitem{Noack2011book}
Bernd~R Noack, Marek Morzynski, and Gilead Tadmor.
\newblock {\em Reduced-order modelling for flow control}, volume 528.
\newblock Springer Science \& Business Media, 2011.

\bibitem{carlberg2011efficient}
Kevin Carlberg, Charbel Bou-Mosleh, and Charbel Farhat.
\newblock Efficient non-linear model reduction via a least-squares
  {P}etrov--{G}alerkin projection and compressive tensor approximations.
\newblock {\em International Journal for Numerical Methods in Engineering},
  86(2):155--181, 2011.

\bibitem{carlberg2017galerkin}
Kevin Carlberg, Matthew Barone, and Harbir Antil.
\newblock Galerkin v. least-squares {P}etrov--{G}alerkin projection in
  nonlinear model reduction.
\newblock {\em Journal of Computational Physics}, 330:693--734, 2017.

\bibitem{Noack2003jfm}
B.~R. Noack, K.~Afanasiev, M.~Morzynski, G.~Tadmor, and F.~Thiele.
\newblock A hierarchy of low-dimensional models for the transient and
  post-transient cylinder wake.
\newblock {\em Journal of Fluid Mechanics}, 497:335--363, 2003.

\bibitem{tadmor2011reduced}
Gilead Tadmor, Oliver Lehmann, Bernd~R Noack, Laurent Cordier, Jo{\"e}l
  Delville, Jean-Paul Bonnet, and Marek Morzy{\'n}ski.
\newblock Reduced-order models for closed-loop wake control.
\newblock {\em Philosophical Transactions of the Royal Society A: Mathematical,
  Physical and Engineering Sciences}, 369(1940):1513--1524, 2011.

\bibitem{morzynski2007continuous}
Marek Morzy{\'n}ski, Witold Stankiewicz, Bernd~R Noack, Rudibert King, Frank
  Thiele, and Gilead Tadmor.
\newblock Continuous mode interpolation for control-oriented models of fluid
  flow.
\newblock In {\em Active Flow Control}, pages 260--278. Springer, 2007.

\bibitem{Rowley2009jfm}
C.~W. Rowley, I.\ Mezi\'c, S.\ Bagheri, P.\ Schlatter, and D.S. Henningson.
\newblock Spectral analysis of nonlinear flows.
\newblock {\em J.\ Fluid Mech.}, 645:115--127, 2009.

\bibitem{Schmid2010jfm}
P.~J. Schmid.
\newblock Dynamic mode decomposition of numerical and experimental data.
\newblock {\em Journal of Fluid Mechanics}, 656:5--28, August 2010.

\bibitem{Tu2014jcd}
J.~H. Tu, C.~W. Rowley, D.~M. Luchtenburg, S.~L. Brunton, and J.~N. Kutz.
\newblock On dynamic mode decomposition: theory and applications.
\newblock {\em Journal of Computational Dynamics}, 1(2):391--421, 2014.

\bibitem{Kutz2016book}
J.~N. Kutz, S.~L. Brunton, B.~W. Brunton, and J.~L. Proctor.
\newblock {\em Dynamic Mode Decomposition: Data-Driven Modeling of Complex
  Systems}.
\newblock SIAM, 2016.

\bibitem{kerswell}
Chris C.~T. Pringle and Rich~R. Kerswell.
\newblock Asymmetric, helical, and mirror-symmetric traveling waves in pipe
  flow.
\newblock {\em Phys. Rev. Lett.}, 99:074502, Aug 2007.

\bibitem{bagheri2009jfm}
Shervin Bagheri, DS~Henningson, J~Hoepffner, and PJ~Schmid.
\newblock Input-output analysis and control design applied to a linear model of
  spatially developing flows.
\newblock {\em Applied Mechanics Reviews}, 62(2), 2009.

\bibitem{rowley2004physd}
Clarence~W Rowley, Tim Colonius, and Richard~M Murray.
\newblock Model reduction for compressible flows using {POD} and {G}alerkin
  projection.
\newblock {\em Physica D: Nonlinear Phenomena}, 189(1-2):115--129, 2004.

\bibitem{Reiss2018jsc}
Julius Reiss, Philipp Schulze, J{\"o}rn Sesterhenn, and Volker Mehrmann.
\newblock The shifted proper orthogonal decomposition: A mode decomposition for
  multiple transport phenomena.
\newblock {\em SIAM Journal on Scientific Computing}, 40(3):A1322--A1344, 2018.

\bibitem{balajewicz2013jfm}
Maciej~J Balajewicz, Earl~H Dowell, and Bernd~R Noack.
\newblock Low-dimensional modelling of high-reynolds-number shear flows
  incorporating constraints from the navier--stokes equation.
\newblock {\em Journal of Fluid Mechanics}, 729:285--308, 2013.

\bibitem{kirby1992reconstructing}
Michael Kirby and Dieter Armbruster.
\newblock Reconstructing phase space from {PDE} simulations.
\newblock {\em Zeitschrift f{\"u}r angewandte Mathematik und Physik ZAMP},
  43(6):999--1022, 1992.

\bibitem{Rowley2000physd}
Clarence~W Rowley and Jerrold~E Marsden.
\newblock Reconstruction equations and the {K}arhunen--{L}o{\`e}ve expansion
  for systems with symmetry.
\newblock {\em Physica D: Nonlinear Phenomena}, 142(1-2):1--19, 2000.

\bibitem{Rim2018juq}
Donsub Rim, Scott Moe, and Randall~J LeVeque.
\newblock Transport reversal for model reduction of hyperbolic partial
  differential equations.
\newblock {\em SIAM/ASA Journal on Uncertainty Quantification}, 6(1):118--150,
  2018.

\bibitem{Iollo2014pre}
Angelo Iollo and Damiano Lombardi.
\newblock Advection modes by optimal mass transfer.
\newblock {\em Physical Review E}, 89(2):022923, 2014.

\bibitem{Lucia2001cfd}
David Lucia, Paul King, Mark Oxley, and Philip Beran.
\newblock Reduced order modeling for a one-dimensional nozzle flow with moving
  shocks.
\newblock In {\em 15th AIAA Computational Fluid Dynamics Conference}, page
  2602, 2001.

\bibitem{Cagniart2019book}
Nicolas Cagniart, Yvon Maday, and Benjamin Stamm.
\newblock Model order reduction for problems with large convection effects.
\newblock In {\em Contributions to Partial Differential Equations and
  Applications}, pages 131--150. Springer, 2019.

\bibitem{Mojgani2017arxiv}
Rambod Mojgani and Maciej Balajewicz.
\newblock Lagrangian basis method for dimensionality reduction of convection
  dominated nonlinear flows.
\newblock {\em arXiv preprint arXiv:1701.04343}, 2017.

\bibitem{Fedele2015jfm}
Francesco Fedele, Ozeair Abessi, and Philip~J Roberts.
\newblock Symmetry reduction of turbulent pipe flows.
\newblock {\em Journal of Fluid Mechanics}, 779:390--410, 2015.

\bibitem{Gerbeau2014jcp}
Jean-Fr{\'e}d{\'e}ric Gerbeau and Damiano Lombardi.
\newblock Approximated {L}ax pairs for the reduced order integration of
  nonlinear evolution equations.
\newblock {\em Journal of Computational Physics}, 265:246--269, 2014.

\bibitem{rasche2017rapid}
Christoph Rasche.
\newblock Rapid contour detection for image classification.
\newblock {\em IET Image Processing}, 12(4):532--538, 2017.

\bibitem{Brunton2016pnas}
Steven~L Brunton, Joshua~L Proctor, and J~Nathan Kutz.
\newblock Discovering governing equations from data by sparse identification of
  nonlinear dynamical systems.
\newblock {\em Proceedings of the National Academy of Sciences}, page
  201517384, 2016.

\bibitem{tibshirani1996regression}
Robert Tibshirani.
\newblock Regression shrinkage and selection via the lasso.
\newblock {\em Journal of the Royal Statistical Society: Series B
  (Methodological)}, 58(1):267--288, 1996.

\bibitem{zheng2018ieee}
Peng Zheng, Travis Askham, Steven~L Brunton, J~Nathan Kutz, and Aleksandr~Y
  Aravkin.
\newblock A unified framework for sparse relaxed regularized regression: {SR3}.
\newblock {\em IEEE Acess}, 7:1404--1423, 2018.

\bibitem{witten2016data}
Ian~H Witten, Eibe Frank, Mark~A Hall, and Christopher~J Pal.
\newblock {\em Data Mining: Practical machine learning tools and techniques}.
\newblock Morgan Kaufmann, 2016.

\bibitem{ng2002spectral}
Andrew~Y Ng, Michael~I Jordan, and Yair Weiss.
\newblock On spectral clustering: Analysis and an algorithm.
\newblock In {\em Advances in neural information processing systems}, pages
  849--856, 2002.

\bibitem{ablowitz2011nonlinear}
Mark~J Ablowitz.
\newblock {\em Nonlinear dispersive waves: asymptotic analysis and solitons},
  volume~47.
\newblock Cambridge University Press, 2011.

\bibitem{kassam2005fourth}
Aly-Khan Kassam and Lloyd~N Trefethen.
\newblock Fourth-order time-stepping for stiff {PDE}s.
\newblock {\em SIAM Journal on Scientific Computing}, 26(4):1214--1233, 2005.

\bibitem{Candes:2011}
E.~J. Cand\`es, X.~Li, Y.~Ma, and J.~Wright.
\newblock Robust principal component analysis?
\newblock {\em Journal of the ACM}, 58(3):11--1--11--37, 2011.

\end{thebibliography}

\end{document}